\documentclass{article}

\usepackage{arxiv}

\usepackage{url}            
\usepackage{algorithmic}
\usepackage{amsmath}
\usepackage{amsfonts}
\usepackage[colorinlistoftodos]{todonotes}
\usepackage{float}
\usepackage{url}            

\usepackage{parskip}  

\DeclareFontFamily{U}{wncy}{}
\DeclareFontShape{U}{wncy}{m}{n}{<->wncyr10}{}
\DeclareSymbolFont{mcy}{U}{wncy}{m}{n}
\DeclareMathSymbol{\comb}{\mathord}{mcy}{"58} 

\usepackage{accents}
\newlength{\dhatheight}

\title{Utilizing the Wavelet Transform's Structure \\ in Compressed Sensing}

\author{
  Nicholas Dwork\thanks{www.nicholasdwork.com, nicholas.dwork@ucsf.edu} \\
  Department of Radiology and Biomedical Imaging \\
  University of California in San Francisco
    \And
  Daniel O'Connor \\
  Department of Mathematics and Statistics \\
  University of San Francisco
    \And
  Corey A. Baron \\
  Robarts Research Institute \\
  The University of Western Ontario
    \And
  Ethan M. I. Johnson \\
  Department of Biomedical Engineering \\
  Northwestern University
    \And
  Adam B. Kerr \\
  Stanford Center for Cognitive and Neurobiological Imaging \\
  Stanford University
    \And
  John M. Pauly \\
  Department of Electrical Engineering \\
  Stanford University
    \And
  Peder E. Z. Larson \\
  Department of Radiology and Biomedical Imaging \\
  University of California in San Francisco
}

\begin{document}
\maketitle

\begin{abstract}
  Compressed sensing has empowered quality image reconstruction with fewer data samples than previously thought possible.  These techniques rely on a sparsifying linear transformation.
  The Daubechies\\ wavelet transform is commonly used for this purpose.  In this work, we take advantage of the structure of this wavelet transform and identify an affine transformation that increases the sparsity of the result.  After inclusion of this affine transformation, we modify the resulting optimization problem to comply with the form of the Basis Pursuit Denoising problem.  Finally, we show theoretically that this yields a lower bound on the error of the reconstruction and present results where solving this modified problem yields images of higher quality for the same sampling patterns using both magnetic resonance and optical images.\footnote{Early versions of this work were presented at the 2020 Joint Mathematics Meeting and at the 2020 Data Sampling workshop of the International Society of Magnetic Resonance in Medicine.}.
\end{abstract}


\section{Introduction}
\label{sec:intro}

Reducing the number of data samples required to generate a quality image is often beneficial.  Compressed sensing \cite{candes2008introduction} has been a remarkable advancement to this end for several imaging systems including Magnetic Resonance Imaging (MRI) \cite{cheng2015free,lustig2007sparse}, computed tomography (CT) \cite{choi2010compressed}, radio astronomy (RA) \cite{li2011application,wiaux2009compressed}, and optical imaging \cite{oike2012cmos}.  Many imaging applications (e.g. MRI, CT, RA) collect samples in the Fourier domain.  Conventionally, the high quality of compressed sensing relied on a sparsifying transform such as the wavelet transform \cite{candes2006robust} or the gradient operator \cite{poon2015role}.  More recently, learning techniques that use a training set of data to (either explicitly or implicitly) determine a basis for signal reconstruction have been created \cite{zhang2018ista,zhu2018image,sandino2020compressed}.  However, learning methods can suffer from over-fitting to the training data, which may take the form of extreme sensitivity to small noise, hallucinations, or an inability to accurately represent the image \cite{antun2019instabilities,maier2019gentle}.  Wavelets provide an orthogonal computationally efficient transform \cite{folberth2016efficient} that, when combined with Fourier sensing, satisfy asymptotic incoherence and multi-level sparsity required for accurate reconstruction \cite{adcock2017breaking}.  Due to these advantages, the wavelet transform continues to be used in compressed sensing applications \cite{blunck2020compressed,datta2019group,huang2020error,baron2018rapid,dai2020ultrasonic,xu2019echo,zhang2019improving}.

The vast majority of compressed sensing research has considered a general linear transformation as the sparsifying transform.
The Discrete Daubechies wavelet transform (DDWT) is an effective choice for compressed sensing \cite{majumdar2012choice}.  One might hope that when this specific transform is used, the reconstruction algorithm could take advantage of its properties to improve the quality of compressed sensing results.  That is the subject of this work.  We identify an affine transformation, based on the DDWT, that increases sparsity.  And we present a method to adapt this transformation into the existing compressed sensing framework.

\section{Theory}
\label{sec:theory}

With compressed sensing of noisy imaging systems, one solves the following Basis Pursuit Denoising (BPD) problem:
\begin{equation}
  \begin{aligned}
    \underset{x\in\mathbb{C}^N}{\text{minimize}} &\hspace{0.5em} \|x\|_1 \hspace{1em} \\
    \text{subject to} &\hspace{0.5em} (1/2)\| A\,x - b \|_2^2 \leq \epsilon,
  \end{aligned}
  \label{eq:basisPursuit}
\end{equation}
where $A$ is the system matrix, $b$ is the data vector, $2\epsilon\geq 0$ is a bound on the noise power \cite{becker2011nesta}, and $\|\cdot\|_p$ is the $L_p$ norm.  This problem can be solved efficiently with standard algorithms (e.g. the Fast Iterative Shrinkage Threshold Algorithm - FISTA \cite{beck2009fast,scheinberg2014fast}).

For many imaging systems, including those discussed in the introduction, $A=MFW^{-1}$, where $M$ is a sampling mask, $F$ is the discrete Fourier transform, $W$ is the DDWT, and $x$ is the vector of wavelet coefficients.  Once \eqref{eq:basisPursuit} is solved, the image can be recovered with $y=W^T\,x$.  This system matrix $A$ satisfies RIPL \cite{bastounis2017absence} which guarantees that $x$ can be recovered robustly \cite{bastounis2017global} and that it solves the corresponding sparse signal recovery problem \cite{chen2001atomic,chen1994basis}.

\begin{figure}[h]
  \centering{}
  \includegraphics[width=0.7\linewidth]{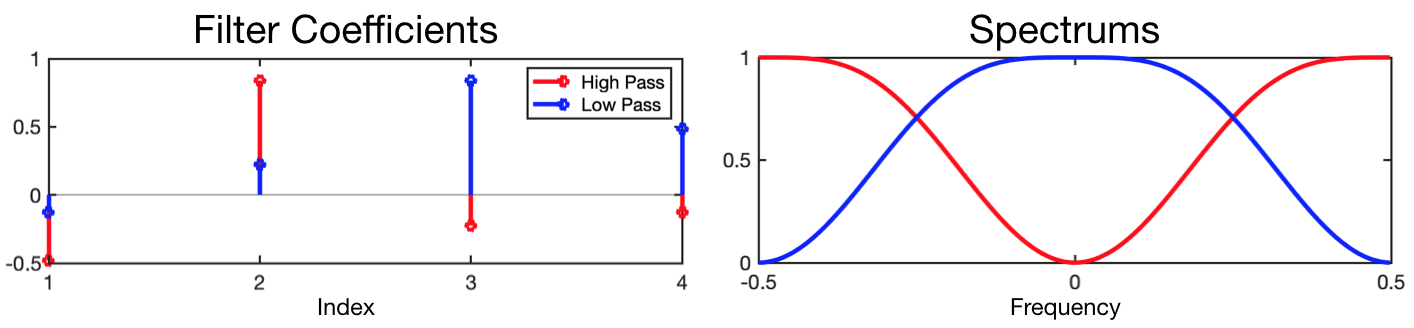}
  \caption{ \label{fig:ddwt} Discrete Daubechies-4 Wavelet Transform.  (Top) The coefficients of the FIR filters.  (Bottom) The spectrum of the same filters (sinc interpolated to 512 elements).}
\end{figure}

\begin{figure*}
  \centering{}
  \includegraphics[width=0.7\linewidth]{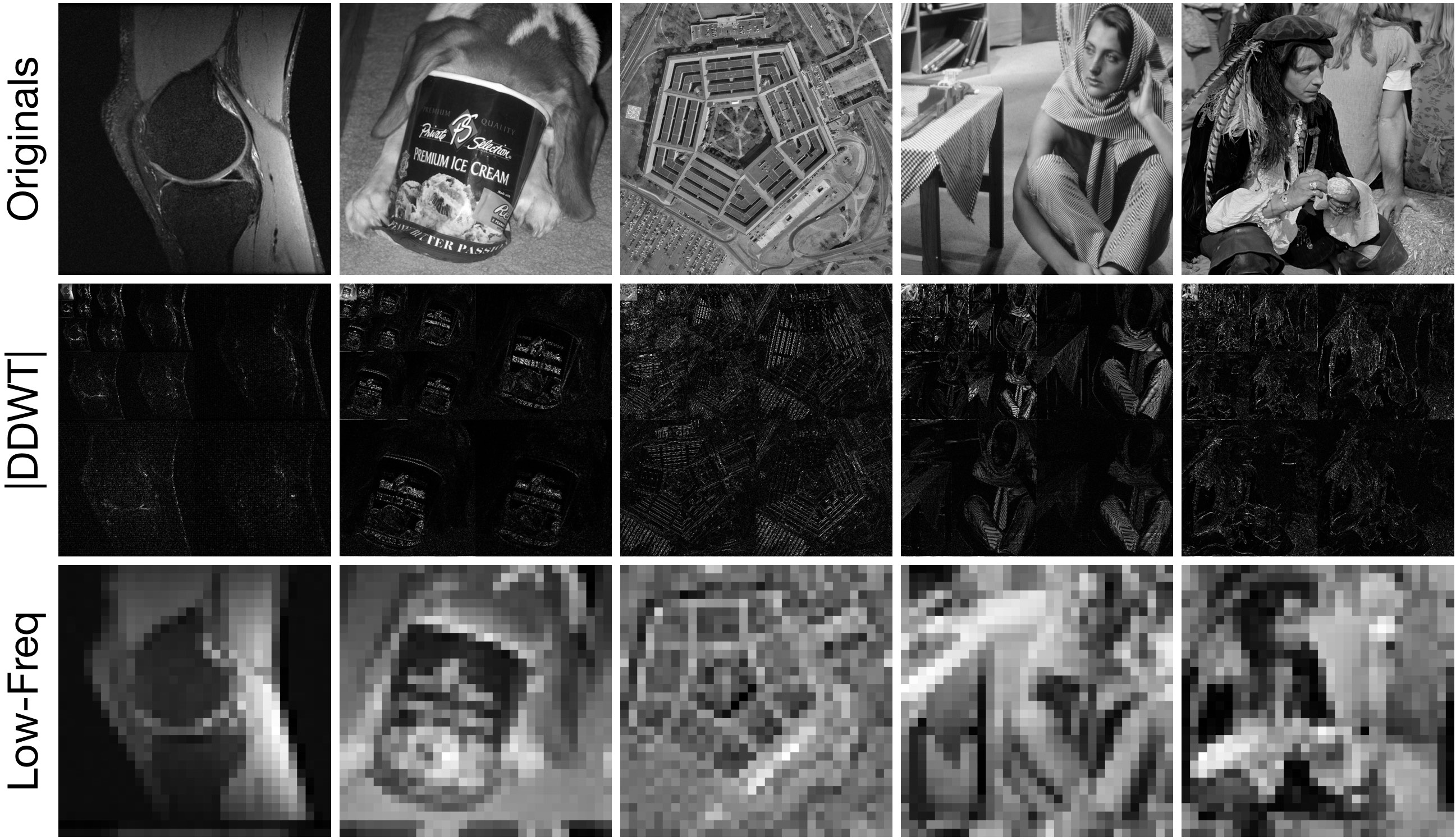}
  \caption{ \label{fig:wavTransforms} Wavelet Transforms of several different images.  (Top) the original images, (Center) The magnitude of the Daubechies-4 wavelet transforms, (Bottom) the lowest-frequency bin of the wavelet transforms.  Notably, the low frequency bin of the transform is not sparse (almost all values are non-zero). }
\end{figure*}

The DDWT consists of two perfect reconstruction finite impulse response (FIR) filters: a low pass and a high pass filter.  There are different types of DDWT, which correspond to different orders of the filters (different numbers of filter coefficients).   The coefficients of the DDWT-4 and their spectrums are shown in Fig. \ref{fig:ddwt}.  The process of applying a wavelet transform is to apply each filter and downsample by $2$ (and then concatenate the results).  We label the response of each filter/downsample as a \textit{bin}.  This process may be applied recursively to each resulting bin.  In the images of Fig. \ref{fig:wavTransforms}, the recursion was applied four times only to the lowest frequency bin.  After transforming, the lowest frequency bin is downsampled by $16$.

Since the composition of FIR low-pass filters is an FIR low-pass filter, the lowest frequency bin in the transform domain is a low-pass filtered (and downsampled) image.  The DDWT low-pass filters are not ideal filters (meaning that they have a non-zero transition region and that their support extends beyond the cutoff frequency); therefore, the lowest frequency bin of the transform also includes aliasing.  Since the other bins of the Wavelet transform of natural images are sparse, by the sifting property of convolution, the aliasing artifacts present in the lowest frequency bin must also be sparse.

To estimate the lowest frequency bin, since it is not typically sparse for natural images, there is no advantage to utilizing the theory of compressed sensing.  Instead, we can rely on the Nyquist-Shannon sampling theorem \cite{bracewell1995two} and collect evenly spaced samples at twice the cutoff frequency.
This technique, fully sampling a region centered on the $0$ frequency in the Fourier domain, was developed as a heuristic and has been used extensively in compressed sensing \cite{uecker2014espirit,vasanawala2011practical,saranathan2012differential,levine20173d}.  The size and shape of the fully sampled region, though, had not been theoretically justified in these techniques.  For example, the SAKE method synthesizes a fully sampled square region of size $80\times 80$ to reconstruct an image of size $200\times 200$ \cite{shin2014calibrationless}.  The DISCO method uses a fully sampled spherical region \cite{saranathan2012differential,levine20173d}.  By only collecting the number of samples required to satisfy the Nyquist-Shannon theorem, the total number of samples is reduced over these heuristic techniques.

Let $r$ denote the number of recursion levels of the DDWT.  Then the size of the lowest frequency bin, after downsampling, would be $(M/2^r) \times (N/2^r)$ pixels$^2$, where the image is size $M\times N$.  Thus, its resolution would be $(2^r/M) \times (2^r/N)$.  According to the Nyquist-Shannon theorem, this image can be reconstructed accurately if a rectangular region of size $(2^r/M) \times (2^r/N)$ with evenly spaced samples centered on the $0$ frequency is collected; we denote this subset of samples as the Fully Sampled Region (FSR).

\section{Methods}
\label{sec:methods}

Our approach will be to first reconstruct a low-frequency image and then correct this estimate with high frequency information.
One could reconstruct a low frequency image simply by performing an Inverse DFT on the data collected in the FSR.  However, doing so leads to ringing (Gibbs phenomenon), which increases the energy in the high frequency bins of the wavelet transform of the image.  Instead, we apply a separable Kaiser-Bessel window \cite{kaiser1974nonrecursive,oppenheim2014discrete} with a parameter of $4$.

Let $\hat{y}_L=F^\ast\,K_B\,M_L\,b$, where $F^\ast$ is the adjoint of the unitary $F$, and $M_L$ and $K_B$ are diagonal matrices representing the mask of the fully sampled region and the Kaiser-Bessel kernel, respectively.
Then we wish to estimate $y$ by solving the following problem:
\begin{equation}
  \begin{aligned}
    \underset{y\in\mathbb{C}^N}{\text{minimize}} &\hspace{0.5em} \| W \, ( y - \hat{y}_L ) \|_1  \\
    \text{subject to} &\hspace{0.5em} (1/2) \, \|M\,F\,y - b\|_2^2 < \epsilon.
  \end{aligned}
  \label{eq:moreSparseImageRecon}
\end{equation}

\begin{figure}[h]
  \centering{}
  \includegraphics[width=0.95\linewidth]{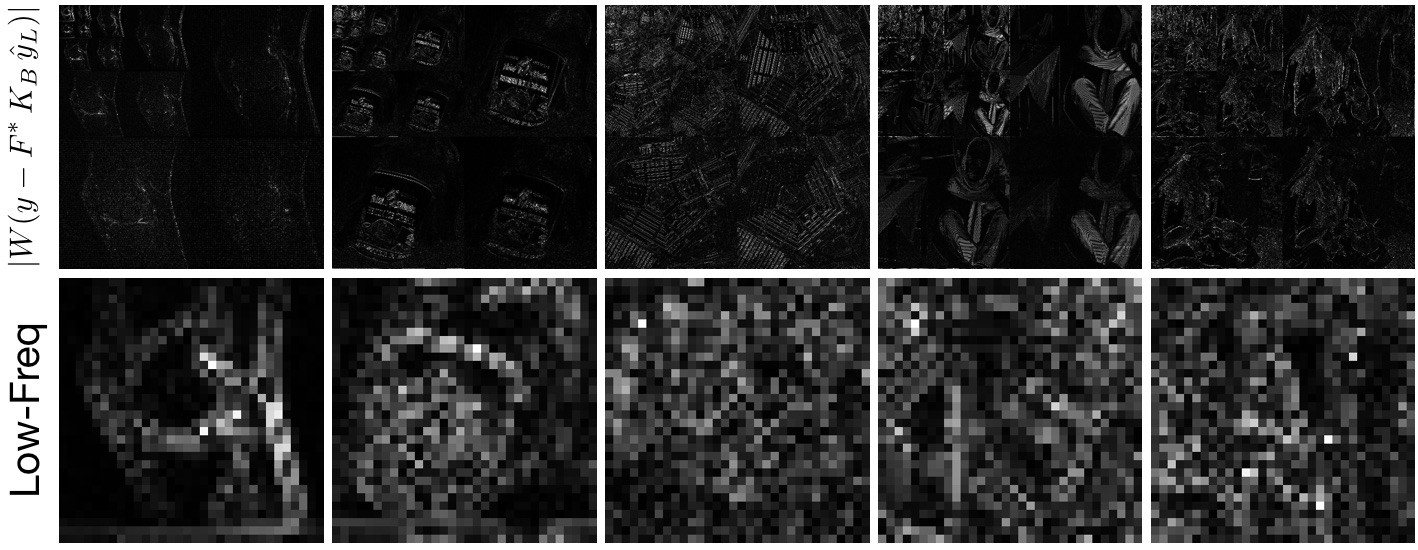}
  \caption{ \label{fig:lowFreqWavTransforms} Wavelet transforms of images after subtracting the low-frequency reconstruction from the image.  (Top) Wavelet transform with $r=4$.  (Bottom) lowest-frequency bin of the Wavelet transform.  The lowest-frequency bin has many more values approximately equal to $0$ than those of Fig. \ref{fig:wavTransforms}.  }
\end{figure}

The key insight is that $z=W \, ( y - \hat{y}_L )$ will be more sparse (have more values approximately equal to $0$) than $x$ for the reasons presented in section \ref{sec:theory}.  Figure \ref{fig:lowFreqWavTransforms} shows $z$ for the images of Fig. \ref{fig:wavTransforms}.  Indeed, they are now better approximated by a sparse representation.

With this definition of $z$, \eqref{eq:moreSparseImageRecon} is equivalent to
\begin{equation}
  \begin{aligned}
    \underset{z\in\mathbb{C}^N}{\text{minimize}} &\hspace{0.5em} \|z\|_1 \hspace{0.5em} \\
    \text{subject to} &\hspace{0.5em} (1/2) \, \| A\,z - \beta \|_2^2 < \epsilon,
  \end{aligned}
  \label{eq:moreSparseBasisPursuit}
\end{equation}
where $\beta=b - M \, F \, \hat{y}_L$.  After solving \eqref{eq:moreSparseBasisPursuit} for $z^\star$, the image can be reconstructed with $y^\star = W^{-1} \, z^\star + \hat{y}_L$.  We call this the More Sparse Basis Pursuit Denoising (MSBPD) solution.  Problem \eqref{eq:moreSparseBasisPursuit} is also a basis pursuit denoising problem; the difference between it and \eqref{eq:basisPursuit} is the data vector is replaced with $\beta$.  Since $A$ satisfies RIPL \cite{li2019compressed}, the bound on the error provided in \cite{bastounis2014absence} applies.  Since the vector $z$ has more elements with value closer to $0$, the error that results from solving \eqref{eq:moreSparseBasisPursuit} has a smaller bound than that of solving \eqref{eq:basisPursuit}.

\section{Results}
\label{sec:results}

\begin{figure}[h]
  \centering{}
  \includegraphics[width=0.95\linewidth]{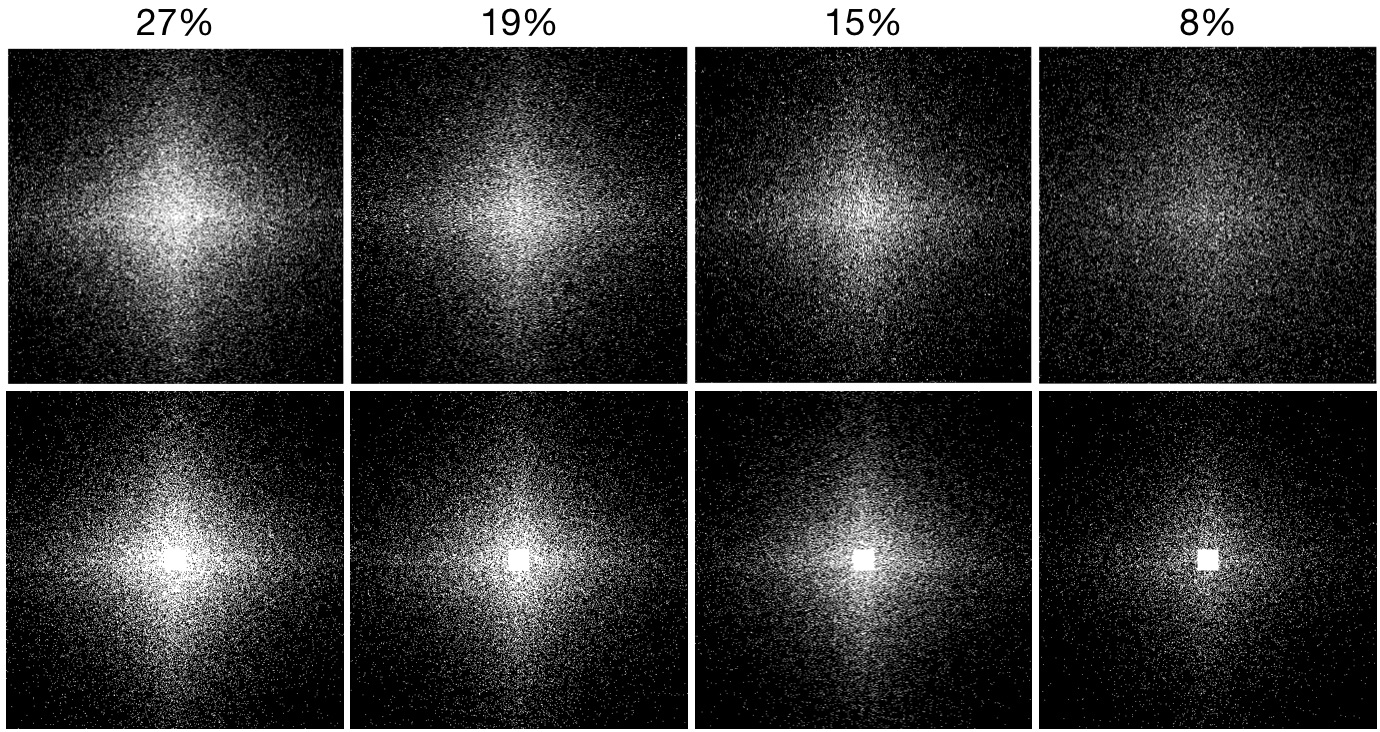}
  \caption{ \label{fig:samplingPatterns} Variable density Fourier sampling patterns generated according to a separable Laplacian distribution.  The top / bottom rows show the sampling patterns without / with the Fully sampled Region (FSR), respectively.  The number of variable density samples is reduced when the FSR is included to maintain the sampling percentage. }
\end{figure}

All experiments in this study were performed with images of size $512\times 512$.  All optimization problems were solved using FISTA with line search \cite{scheinberg2014fast,beck2009fast} run for $100$ iterations.  The wavelet transform applied was the DDWT-4, recursively applied to only the lowest-frequency bin $r=4$ times.
Figure \ref{fig:samplingPatterns} shows the sampling patterns used in the experiments with various sampling percentages with and without the fully sampled center region.  The sampling patterns are realizations of a random separable Laplacian distribution \cite{fang2016high} with a standard deviation of approximately 20\%.  This pattern has a higher probability of sampling higher frequencies than a Gaussian distribution with the same standard deviation.  This likely better matches the actual power spectral density of natural images and is, therefore, a more optimal sampling pattern for a compressed sensing reconstruction \cite{adcock2014quest}.  For the sampling patterns with the fully sampled region, fewer samples from the distribution were included in order to retain approximately the same number of total samples.
The value of $\epsilon$ was chosen independently for each reconstruction by conducting an exhaustive search to find the value that minimized the relative error $e = \| y^\star - y \|_2 / \| y \|_2 $.

\begin{figure*}
  \centering{}
  \includegraphics[width=0.7\linewidth]{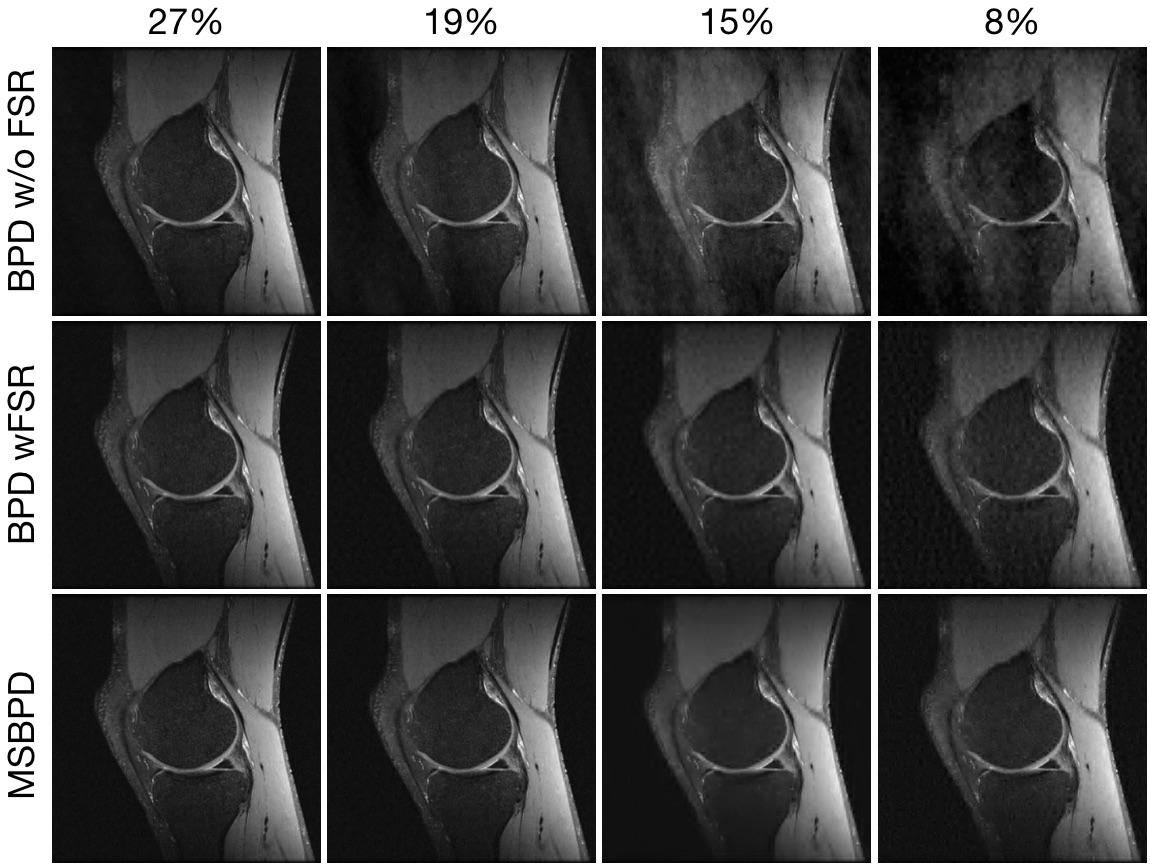}
  \caption{ \label{fig:recons_nSamples} Reconstructions of MR images of the the knee with different sampling percentages without and with the Fully Sampled Region (FSR). }
\end{figure*}

Figure \ref{fig:recons_nSamples} shows reconstructions using BPD and MSBPD.  As can be observed in Fig. \ref{fig:recons_nSamples}, MSBPD offers the highest quality reconstruction for all sampling percentages.  The improvements become more noticeable as the sampling percentage is reduced.

Table \ref{tbl:reconErrorsNumSamples} shows the relative errors for the reconstructions in Fig. \ref{fig:recons_nSamples}.  In all cases, MSBPD yields the lowest relative error.  The most significant improvement comes by including the FSR in the samples; a minor gain is attained by using MSBPD over BPD on this sampling pattern.  The improvement is more pronounced as the sampling percentage decreases.

\begin{table}[ht]
  \centering
  \caption{Relative errors for reconstructions of knee image from Fig. \protect\ref{fig:wavTransforms} with different sampling percentages.}
  \begin{tabular}{||l c c c c||} 
  \hline
  Sampling Percentage & 27\% & 19\% & 15\% & 8\% \\
  \hline
  BPD       & 0.059 & 0.070 & 0.084 & 0.143  \\
  BPD w/FSR & 0.058 & 0.068 & 0.076 & 0.113  \\
  MSBPD     & 0.056 & 0.065 & 0.071 & 0.093 \\
  \hline
  \end{tabular}
  \label{tbl:reconErrorsNumSamples}
\end{table}

Table \ref{tbl:ssimNumSamples} shows the Structural Similarity metric (SSIM) comparing the reconstructed images to the originals.  This quantification shows the same trends as the relative errors.

\begin{table}[ht]
  \centering
  \caption{SSIM values for reconstructions of knee image from Fig. \protect\ref{fig:wavTransforms} with different sampling percentages.}
  \begin{tabular}{||l c c c c||} 
  \hline
  Sampling Percentage & 27\% & 19\% & 15\% & 8\% \\
  \hline
  BPD       & 0.996 & 0.995 & 0.992 & 0.977  \\
  BPD w/FSR & 0.996 & 0.995 & 0.994 & 0.986  \\
  MSBPD     & 0.997 & 0.995 & 0.995 & 0.990 \\
  \hline
  \end{tabular}
  \label{tbl:ssimNumSamples}
\end{table}

\begin{figure}[h]
  \centering{}
  \includegraphics[width=0.95\linewidth]{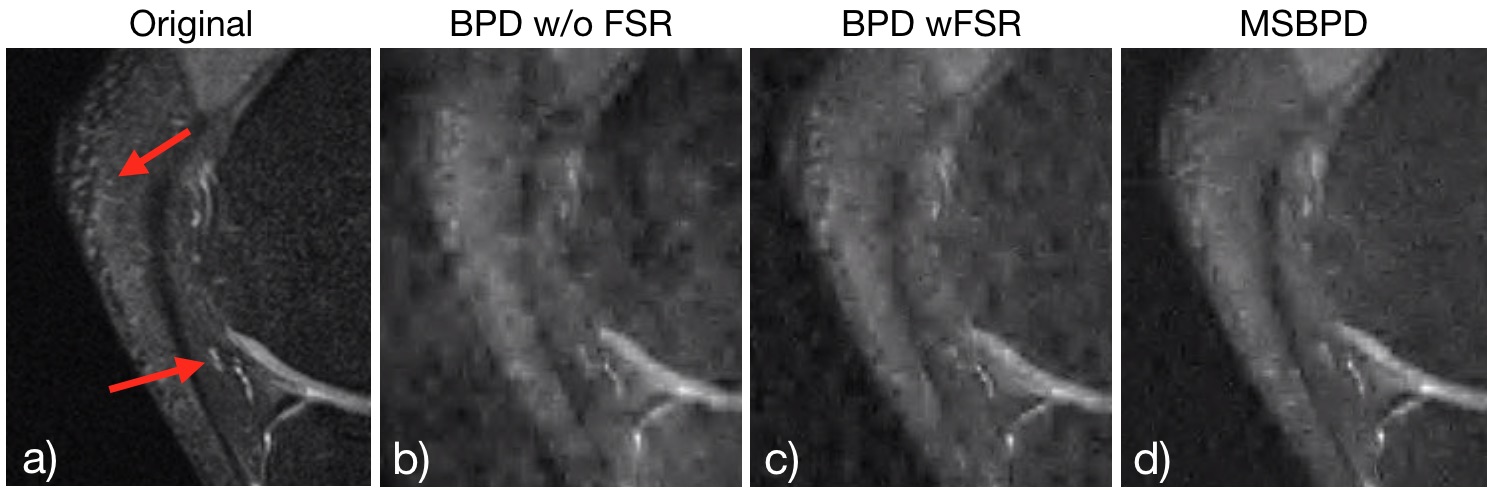}
  \caption{ \label{fig:zoomInKnee} Zoom in to reconstructions of MRI data of knee from 8\% of the data.  (a) shows the original image;  (b) shows the reconstruction using BPD with variable density data;  (c) shows the reconstruction using BPD with a fully sampled center region; and (d) shows reconstruction with MSBPD.  The red arrows point to regions in the image where the improvement in quality of MSBPD over the other algorithms is very apparent.}
\end{figure}

In Fig. \ref{fig:zoomInKnee}, we zoom into the reconstruction of the knee from 8\% of data.  This figure shows fewer wavelet artifacts in the result of MSBPD than exist in the result of BPD with either sampling pattern.

\begin{figure*}
  \centering{}
  \includegraphics[width=0.8\linewidth]{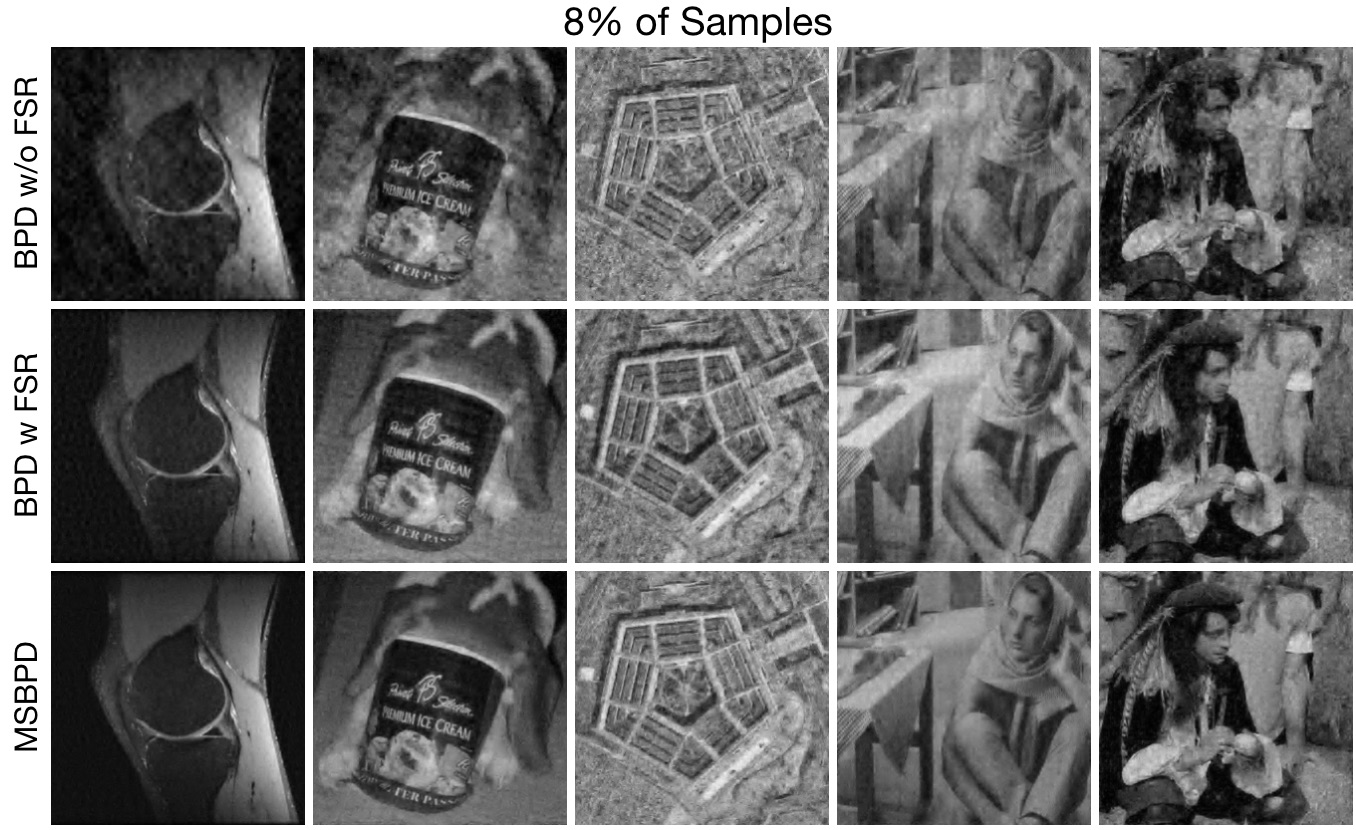}
  \caption{ \label{fig:reconsAllImages} A comparison of BPD (without and with the FSR) to MSBPD with 8\% of the sampling data for all images of Fig. \protect\ref{fig:wavTransforms}.  The quality of reconstruction with MSBPD is improved over the quality of BPD. }
\end{figure*}

\begin{figure}[h]
  \centering{}
  \includegraphics[width=0.95\linewidth]{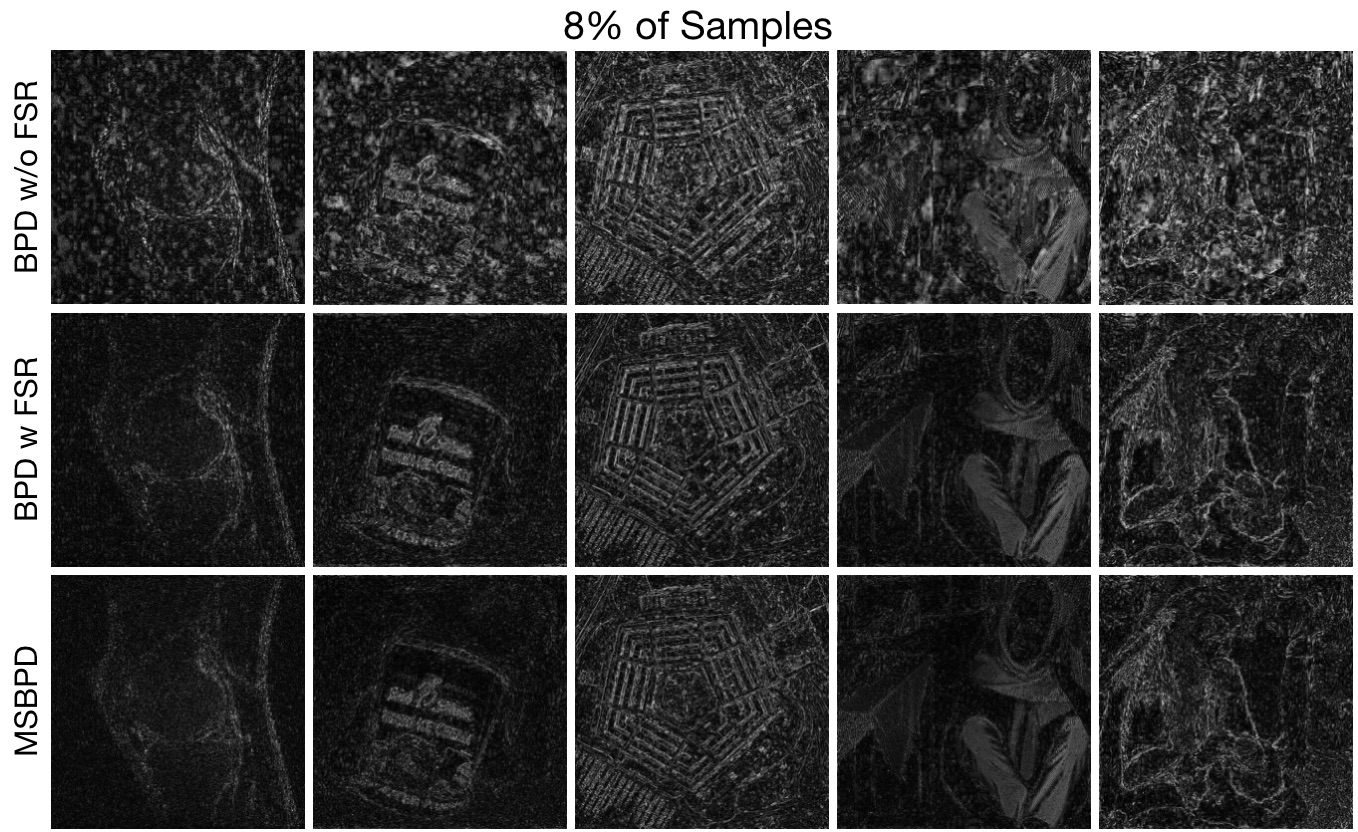}
  \caption{ \label{fig:reconsAllImagesErrors} Magnitudes of the differences between the reconstructions of Fig. \ref{fig:reconsAllImages} and the original images shown in Fig. \protect\ref{fig:wavTransforms}.  There is a significant reduction in the error by including the FSR in the sampling pattern.  An additional reduction in error is realized by using MSBPD. }
\end{figure}

Figure \ref{fig:reconsAllImages} shows the reconstructions with 8\% of the data for BPD and MSBPD on all images of Fig. \ref{fig:wavTransforms}.
Figure \ref{fig:reconsAllImagesErrors} shows the magnitudes of the differences between the reconstructions with 8\% of the data shown in Fig. \ref{fig:reconsAllImages} and the original images.  The errors are significantly reduced by including the FSR.  There is an additional improvement to the reconstruction when using MSBPD.

Table \ref{tbl:reconErrorsAllImages} shows the relative error for each algorithm on each of the five reconstructions of Fig. \ref{fig:reconsAllImages}.  As observed previously, the most significant gain is attained by including the FSR.  A minor additional gain is attained with MSBPD.

\begin{table}[ht]
  \centering
  \caption{Relative error for reconstructions of all images from Fig. \ref{fig:wavTransforms} with 8\% of data.}
  \begin{tabular}{||l c c c c c||} 
  \hline
  Image Index & 1 & 2 & 3 & 4 & 5 \\ [0.5ex] 
  \hline
  BPD       & 0.143 & 0.183 & 0.118 & 0.201 & 0.256 \\
  BPD w/FSR & 0.093 & 0.131 & 0.104 & 0.147 & 0.169 \\
  MSBPD     & 0.093 & 0.113 & 0.103 & 0.143 & 0.159 \\
  \hline
  \end{tabular}
  \label{tbl:reconErrorsAllImages}
\end{table}

Table \ref{tbl:ssimsAllImages} shows the Structural Similarity metric (SSIM) comparing the reconstructed images to the originals.  Once again, this metric shows the same trends as the relative errors.

\begin{table}[ht]
  \centering
  \caption{SSIM values for reconstructions of all images from Fig. \ref{fig:wavTransforms} with 8\% of data.}
  \begin{tabular}{||l c c c c c||} 
  \hline
  Image Index & 1 & 2 & 3 & 4 & 5 \\ [0.5ex] 
  \hline
  BPD       & 0.977 & 0.921 & 0.699 & 0.866 & 0.878 \\
  BPD w/FSR & 0.986 & 0.959 & 0.797 & 0.935 & 0.948 \\
  MSBPD     & 0.990 & 0.970 & 0.806 & 0.939 & 0.954 \\
  \hline
  \end{tabular}
  \label{tbl:ssimsAllImages}
\end{table}

\section{Discussion}
In this work, we have utilized the structure of the Discrete Daubechies Wavelet Transform to improve image reconstruction results based on compressed sensing.  It takes advantage of the prior knowledge that a great deal of images are not sparse in the lowest-frequency bin of the image's wavelet transform.

Note that some images, those that are natively sparse, are indeed sparse in the lowest frequency bin of the DDWT; e.g. images of angiography and some astronomy.  There would be no advantage to using MSBPD over BPD for images of that type, and other types of compressed sensing reconstruction algorithms may be more appropriate \cite{cukur2011signal}.  When it is the case, however, that the image is not sparse in the lowest frequency bin then MSBPD offers reconstruction with improved quality over BPD even when the sampling pattern includes the FSR.

Due to the increased sparsity, an accurate reconstruction may be possible using the greedy Orthogonal Matching Pursuit \cite{pati1993orthogonal,tropp2007signal} algorithm to solve the corresponding sparse signal recovery problem:
\begin{equation*}
  \underset{z\in\mathbb{C}^N}{\text{minimize}} \hspace{0.5em} (1/2) \, \| A \, z - \beta \|_2^2 \hspace{0.5em}
  \text{subject to} \hspace{0.5em} \|z\|_0 \leq \mathcal{S},
\end{equation*}
where $\|\cdot\|_0$ is the $L_0$ penalty and $\mathcal{S}$ limits the number of non-zero elements in the vector $z$.  This could make image reconstruction more computationally efficient.

The DDWT and the DFT are both radially asymmetric transforms; i.e., a rotation operator and the transform operator do not generally commute.  However, in many imaging systems, there is not anything inherently special about the horizontal and vertical directions (e.g. MRI, CT, radio interferometry).  Therefore, it may be possible to reconstruct images of higher quality by collecting a circle of data in the low-frequency region (rather than a square) and use a sparsifying transform that is radially symmetric.  Symmetric wavelets on the sphere \cite{lessig2008soho} may be such a sparsifying transform.  Though this may yield improved quality, the computations required may increase.  This side-effect of this possible improvement should be considered for any given application.

With each recursive application of the Wavelet transform, the average sparsity is reduced \cite{adcock2017breaking}.  Thus, the value of $r$ can be chosen based on the sparsity achievable for the imaging system and subjects of interest.  Additionally, the higher the value of $r$, the more computations are required to implement the wavelet transform.  Both aspects should be considered when determining a value of $r$ for a given application.

In this work, we determined the regularization parameter with an exhaustive search.  This is only appropriate when ground truth is known and not a practical solution.  For some applications, the bound on the noise can be determined with an image of noise; this can be accomplished in MRI, for example, by imaging with an excitation of $0$.  When a calibration image is unavailable, reconstruction may be accomplished by adapting the MSBPD algorithm to an iterative re-weighting algorithm \cite{candes2008enhancing,chartrand2008iteratively,asif2013fast,voronin2015iteratively}.  Alternatively, an automatic parameter tuning algorithm can be used based on existing training data \cite{kunisch2013bilevel} or with properties of the resulting pareto-optimal set \cite{shahdloo2018projection}.  Some of these methods have the consequence of implicitly altering the objective function \cite{candes2008enhancing}, which means that the theorems of compressed sensing no longer hold.  However, they have been shown heuristically to improve image quality in some cases.

Finally, the method presented in this work can be adapted to situations where information is shared amongst several acquisitions.  This can happen with multiple-acquisition MRI \cite{senel2019statistically}, parallel MRI \cite{murphy2012fast}, or synthetic-aperture radar imaging \cite{zhu2015joint} and may involve joint-sparsity regularization \cite{kopanoglu2020simultaneous}.  Moreover, the method can be adapted to deep-learning methods that are regularized with wavelet sparsity or joint sparsity \cite{dar2020prior}.

We leave the investigations of these possible extensions as future work.

\section{Conclusion}
In this work, we have presented the MSBPD algorithm.  This algorithm utilizes the structure and behavior of the DDWT to justify a sampling pattern and to identify a new sparsifying transform that increases sparsity of most natural images.  Since the system matrix satisfies RIPL, this leads to improved results over solving the standard BPD problem of \eqref{eq:basisPursuit}.  In experiments, we compared image quality of reconstructions made with BPD, BPD with the FSR, and MSBPD.  In all cases, the most significant gain was attained by including the FSR in the sampling pattern and MSBPD yielded the lowest relative error.

\section{Acknowledgments}
ND would like to thank the Quantitative Biosciences Institute at UCSF and the American Heart Association as funding sources for this work.  ND is supported by a Postdoctoral Fellowship of the American Heart Association.  ND and PL have been supported by the National Institute of Health’s Grant No. NIH R01 HL136965.


\end{document}